# Critique of "Asynchronous Logic Implementation Based on Factorized DIMS"

P. Balasubramanian
School of Electrical and Electronic Engineering
Nanyang Technological University
50 Nanyang Avenue
Singapore 639798

## Abstract
This paper comments on "Asynchronous Logic Implementation Based on Factorized DIMS" [*Journal of Circuits, Systems, and Computers*, vol. 26, no. 5, 1750087: 1-9, May 2017] with respect to two main problematic issues: i) the gate orphan problem implicit in the factorized DIMS approach discussed in the referenced article which affects its strong-indication, and ii) how the enumeration of product terms to represent the synthesis cost is skewed in the referenced article because the logic expression contains sum of products and also product of sums. It is observed that the referenced article has not provided a general logic synthesis algorithm excepting only an example illustration involving a 3-input AND logic function. The absence of a general logic synthesis algorithm would make it difficult to reproduce the research described in the referenced article. Moreover, the example illustration in the referenced article describes an unsafe logic decomposition which is not suitable for the multi-level synthesis of strong-indication asynchronous circuits. Further, a logic synthesis method which safely decomposes the DIMS solution to synthesize multi-level strong-indication asynchronous circuits is available in the existing literature, which was neither cited nor taken up for comparison in the referenced article, which is another drawback. Subsequently, it is concluded that the referenced article has not advanced existing knowledge in the field but on the contrary, has caused confusions. Hence, in the interest of readers, this paper additionally highlights some important and relevant literature which provide valuable information about robust asynchronous circuit synthesis techniques which employ delay-insensitive codes for data representation and processing and the 4-phase return-to-zero handshake protocol for data communication.

**Keywords:** Logic synthesis; Asynchronous logic; Sum-of-minterms; Delay-insensitive; Disjoint sum of products; Indication

## 1. Introduction
The International Technology Roadmap for Semiconductors (ITRS report) [1] has highlighted variability as one of the grand challenges for electronic design in the nanoelectronics regime. To cope with the crucial design issues such as modularity, parameters uncertainty, variability etc. the asynchronous design method is suggested to be an alternative to the synchronous design method. Especially, asynchronous circuits based on delay-insensitive data codes which utilize a 4-phase handshake protocol are inherently robust as they can absorb variations in process or voltage or temperature parameters [2]. Such asynchronous circuits are basically elastic and are called quasi-delay-insensitive (QDI), which implies the practical implementation of delay-insensitive asynchronous circuits with the only exception and assumption of isochronic forks [3], where isochronic forks form the weakest compromise to delay-insensitivity.

Reference [4] presented a *so-called* strongly indicating asynchronous logic synthesis approach for combinational logic functions by utilizing delay-insensitive dual-rail codes and the 4-phase return-to-zero handshake protocol. The strongly indicating combinational logic synthesis in [4] was labelled as the factorized delay-insensitive minterm synthesis (FDIMS) approach; the original DIMS method was proposed in [5]. Reference [4], which is derived from [5], poses a potential problem of gate orphan(s) with respect to the proposed asynchronous logic synthesis, which could affect the robustness of the synthesized circuit. Note that no gate orphan problem whatsoever exists in [5]. Moreover, there exists safe asynchronous strong- and weak-indication combinational logic synthesis methods [6 – 8] and [17 – 20] which were neither cited nor taken up for comparison in [4]. Further, [4] also embeds a confusion with respect to estimation of the synthesis cost of the resulting solution. In this regard, this paper describes the major problems with [4]. The logic factorization discussed in [4] gives rise to logic expression(s) which contain sum of products and product of sums, which is representative of a mixed logic synthesis [9 – 11]. However, [4] considers only the product terms in the logic expression to give an estimate of the synthesis cost which is erratic.

Further, it is noted that [4] has not provided a generalized logic synthesis algorithm to deduce the FDIMS solution, which is nevertheless unsafe. This is because logic factorization can be performed in different ways [12] [13], and most of the conventional factorization techniques cannot be termed as safe for QDI logic synthesis and/or optimization [7] [14]. Nonetheless, without a generic logic synthesis algorithm, it would be difficult for a researcher to be able to reproduce the research presented in [4] and/or verify the correctness of the results reported.

There are two main problematic issues with [4] which will be described in this report. Firstly, in Section 4 of [4], at the beginning of the first paragraph, the author states "To reduce DIMS complexity and ensure strong-indication, a method called factorized DIMS (i.e. FDIMS) is proposed". It is clear from this statement that the author of [4] intended to achieve the twin objectives of: (i) reducing the complexity of the original DIMS solution through logic factoring, and (ii) ensuring that strong-indication is preserved in the resulting FDIMS solution. Due to the gate orphan(s) problem being imminent in [4], which will be illustrated later in Section 3, the resulting asynchronous circuit synthesized on the basis of [4] becomes less robust than the original DIMS circuit, which is self-defeating. Moreover, this is undesirable and not acceptable since FDIMS has been labelled as a strong-indication synthesis approach in [4] which is not. On the contrary, [7], which is a strongly indicating logic synthesis method that also factorizes the DIMS solution, and which is the closest counterpart to [4], is robust and is guaranteed to be free of gate orphans. The robustness of [7] results from the fact that it employs only safe QDI logic decomposition procedures. However, [6] or [7], which are very relevant to [4], were neither cited nor taken up for comparison in [4]. Besides these, there exists other well-established works in the literature such as [15] [16] which form important references for robust asynchronous logic synthesis.

The remainder of this critique is organized as follows. In the interest of readers, Section 2 gives a background about robust asynchronous circuit design employing delay-insensitive data codes for data representation and processing, and the 4-phase return-to-zero handshake protocol for data communication. Section 3 describes the gate orphan problem in [4] through the example circuit realization given in Section 4 of [4]. How to overcome the gate orphan problem in the example circuit realization given in [4] via a safe QDI logic decomposition is also discussed. Section 4 sheds light on the underestimation of synthesis costs for the combinational logic benchmarks considered in Section 5 of [4]. Finally, the conclusions are given in Section 5.

## 2. Robust Asynchronous Logic Circuits – Background

A background about delay-insensitive data encoding and 4-phase return-to-zero handshaking is first provided followed by a discussion of the types of robust asynchronous circuits.

### 2.1. Delay-Insensitive Dual-Rail Data Encoding and 4-Phase Handshaking

The dual-rail code (a.k.a. 1-of-2 code) is the simplest member of the family of delay-insensitive *m*-of-*n* codes [21]. Among the family of *m*-of-*n* codes, 1-of-*n* codes represent a subset and are called one-hot codes. In a 1-of-*n* code, only 1 out of *n* wires is asserted high i.e. binary 1 to represent a binary data. In fact, the 1-of-*n* coding scheme is said to be unordered [22] since none of the code words forms a subset of another code word. Also, the 1-of-*n* coding scheme is said to be complete [23] if all the *n* unique code words, as per the definition, are utilized to encode the specified binary inputs.

As per the 1-of-2 code, a single-rail binary input say Q, is encoded using two wires as say Q1 and Q0, where Q = 1 is represented by Q1 = 1 and Q0 = 0, and Q = 0 is represented by Q1 = 0 and Q0 = 1. Note that Q1 and Q0 cannot assume 1 concurrently as it is illegal and invalid since the coding scheme will then become unordered. However, Q1 and Q0 can assume 0 concurrently and is referred to as the spacer. Hence, as per the dual-rail code, a valid data is specified by either Q1 or Q0 assuming binary 0 and the other assuming binary 1, and the condition of both Q1 and Q0 assuming binary 0 is called the spacer or null [24].

An asynchronous circuit stage that employs the delay-insensitive dual-rail code for data representation and processing and the 4-phase return-to-zero handshake protocol for data communication is shown in Figure 1. As the name implies, the 4-phase return-to-zero protocol consists of four phases which will be explained with reference to Figure 1 by assuming dual-rail encoded data. Nevertheless, this explanation would be applicable for data represented using any delay-insensitive 1-of-*n* code.

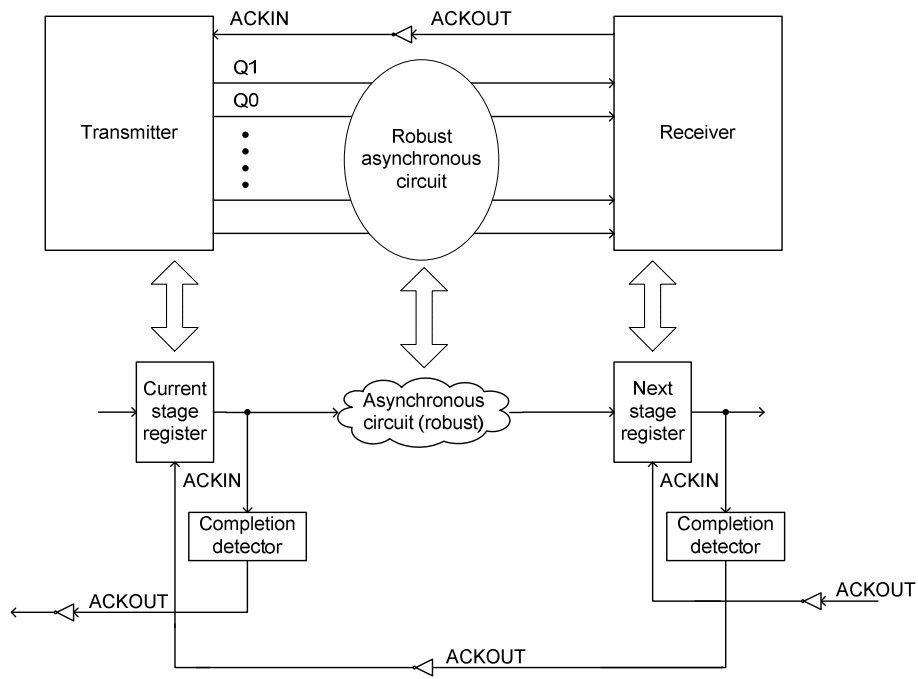

Figure 1. A robust asynchronous circuit stage correlated with the transmitter-receiver analogy

In the first phase, the dual-rail data bus shown in Figure 1 which is specified by (Q1, Q0) etc. is in the spacer state and ACKIN is high i.e. binary 1. The transmitter transmits a code word i.e. valid data and this results in rising signal transitions i.e. binary 0 to 1 on one of the corresponding dual rails of the entire dual-rail data bus. In the second phase, the receiver receives the code word sent, and it drives ACKOUT high. In the third phase, the transmitter waits for ACKIN to go low i.e. binary 0 and then resets the entire dual-rail data bus to the spacer state. Subsequently, in the fourth phase, after an unbounded time duration which is deemed finite and positive, the receiver drives ACKOUT to low i.e. ACKIN becomes high. One data transaction is said to be complete, and the asynchronous circuit stage can commence the next data transaction. Hence the application of input data follows the sequence: valid data-spacer-valid data-spacer, and so forth.

## 2.2. Strong-Indication, Weak-Indication, and Early Output

QDI asynchronous logic circuits are generally categorized as strong-indication [25] [26], weak-indication [25] [27], and early output [28] [29] types. *Indication, at the circuit-level, means providing acknowledgment for the receipt of the primary inputs through the primary outputs. This is accomplished by ensuring that indication is duly provided by the intermediate outputs* [2]. With respect to the asynchronous circuit stage shown in Figure 1, the indication mechanism may be either local or global [30] [31]. The indication mechanism is local if the asynchronous circuit by itself can acknowledge the receipt of all the primary inputs. The indication mechanism is global if the asynchronous circuit stage, overall, indicates the receipt of all the primary inputs in conjunction with the asynchronous circuit present in it. The input-output timing characteristics of strong-indication, weak-indication, and early output asynchronous circuits is illustrated by the representative timing diagram shown in Figure 2.

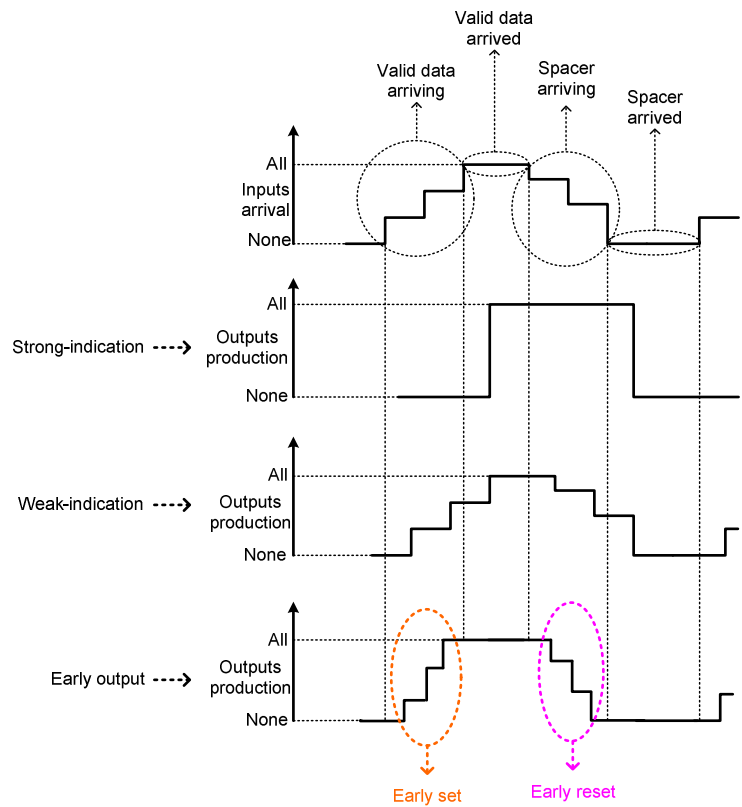

Figure 2. Input-output timing behavior of strong-indication, weak-indication, and early output asynchronous logic circuits

A strong-indication asynchronous circuit starts data processing to produce the required primary outputs only after receiving all the primary inputs whether they are valid data or spacer. A weak-indication asynchronous circuit starts data processing and could produce some of the primary outputs after receiving just a subset of the primary inputs. Nonetheless, the production of at least one primary output is delayed till the last primary input is received. An early output asynchronous circuit could start data processing and produce all the primary outputs after receiving just a subset of the primary inputs. In general, if all the valid primary outputs are produced after receiving valid data on a subset of the primary inputs, the early output asynchronous circuit is said to be of early set type. On the other hand, if the spacer is produced on all the primary outputs after receiving the spacer on a subset of the primary inputs, the early output asynchronous circuit is said to be of early reset type. The early set and early reset behaviors of an early output asynchronous circuit are depicted through the orange and pink ovals in dotted lines in Figure 2. Amongst the timing models, the strong-indication is the most restrictive and the early output is the more relaxed.

## 3. Gate Orphan Problem with the FDIMS Approach of [4]

*In an indicating asynchronous circuit, any transition on the primary inputs are required to propagate monotonically i.e. unidirectionally throughout the entire circuit depth up to the primary outputs with no unacknowledged signal transition on any intermediate gate output* [32]. For proper indication, the signal transitions should either monotonically increase from binary 0 to 1, or monotonically decrease from binary 1 to 0 throughout the entire circuit. For data represented using the dual-rail code and communicated based on the 4-phase return-to-zero handshaking, when valid data are applied the transitions would monotonically increase and for the application of the spacer the transitions would monotonically decrease. Any unacknowledged signal transition on a gate output is termed as *gate orphan*. The issue of gate orphan has been clearly explained in [33] [34], and the gate orphan problem inherent in [4] shall be explained with the same example circuit that was considered for illustration in [4].

Care should be taken to ensure that any logic transformation or optimization performed in an indicating asynchronous circuit conforms to the safe QDI logic decomposition principles. This is because indication and robustness go together in QDI asynchronous circuits. Any arbitrary decomposition of logic gate(s) or C-element(s) might give rise to gate orphan(s) which could potentially affect the robustness of the QDI circuit. Resolving the gate orphan(s) is non-trivial and may require extensive timing analysis [35] and additional timing assumptions which could complicate a circuit's physical implementation. Also, if gate orphans are left unresolved, they might become problematic to a QDI circuit/system operation [19], and may also cause a stall. Thus, gate orphan(s) are to be avoided in indicating asynchronous circuit designs but this has been casually overlooked and conveniently neglected in [4].

Reference [4] shows an example 3-input AND function implemented in asynchronous style which gives rise to gate orphan(s). To maintain consistency with the discussion given in [4], we retain the same dual-rail encoded primary inputs viz. (X31, X30), (X21, X20), (X11, X10), and the dual-rail encoded primary output (f1, f0) for the 3-input AND function. The dual-rail 3-input AND function synthesized based on the DIMS method [5] is shown in Figure 3, which is equivalent to Figure 1 of [4]. In Figure 3, gates C1 to C8 represent 3-input C-elements[1], and OR1 and OR2 are the OR logic gates.

---

[1] The C-element outputs 1 or 0 if all its inputs are 1 or 0 respectively. Otherwise it maintains its existing state. The C-element is represented by the circle with the marking C in the figures.

The logic equations corresponding to the 3-input AND function, expressed based on the DIMS method [5], are given by (1) and (2). Equations (1) and (2) are in the disjoint sum of products (DSOP) form [36 – 38]. In a DSOP form, the logical conjunction of any two product terms yields zero. Only one product term is activated in a DSOP expression for an input pattern. In the equations, product refers to logical conjunction and sum refers to logical disjunction.

f1 = X31X21X11  (1)

f0 = X30X20X10 + X30X20X11 + X30X21X10 + X30X21X11 + X31X20X10 + X31X20X11
   + X31X21X10  (2)

Reference [4] has arbitrarily factorized (2), and has yielded the reduced logic expression (3) for 'f0', which is given below.

f0 = [X30X20 + X30X21 + X31X20 + X31X21]X10 + [X30 + X31]X20X11 + X30X21X11
   = [X30(X20 + X21) + X31(X20 + X21)]X10 + [X30 + X31]X20X11 + X30X21X11
   = [(X30 + X31)(X20 + X21)]X10 + [X30 + X31]X20X11 + X30X21X11  (3)

Since output f1 i.e. (1) contains just one product term, it cannot be minimized further. The circuit synthesized based on (1) and (3), as per the FDIMS approach [4], is shown in Figure 4, which is equivalent to Figure 4 of [4]. In Figure 4, C1, C2, C3, and C4 represent 3-input C-elements, and the three OR gates are labelled as OR1, OR2, and OR3 for the sake of explanation. The node marked 'isf' in Figure 4 is an isochronic fork [3], which implies that a rising or a falling signal transition on isf would be simultaneously transmitted to the respective inputs of the C-elements C2 and C3 in Figure 4.

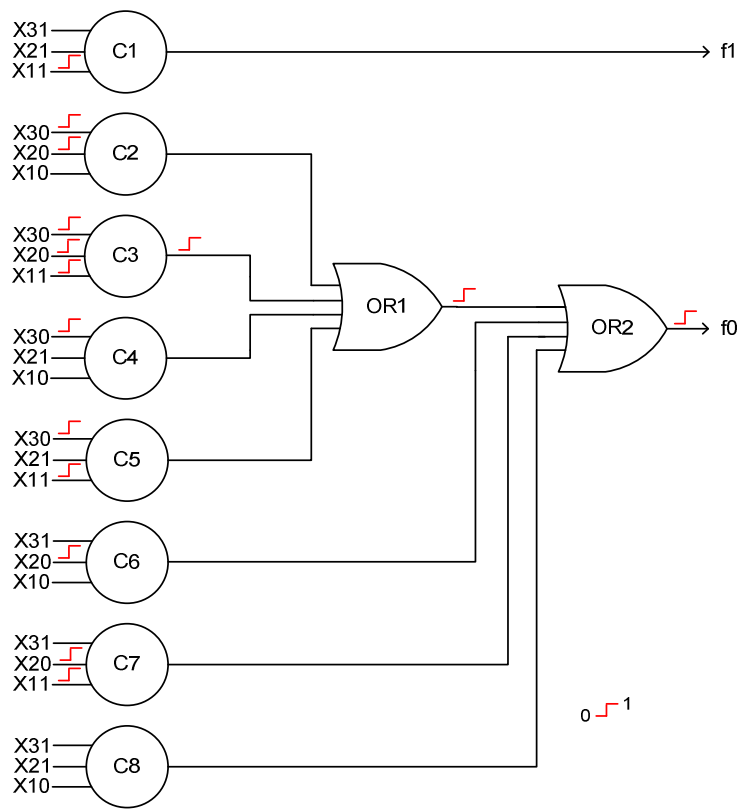

Figure 3. 3-input AND function synthesized according to the DIMS method [5]

In Figure 3, the rising signal transitions on the primary inputs X30, X20 and X11 are acknowledged by a rising signal transition on the output of C3 which is followed by a rising signal transition on the output of OR1 which is subsequently followed by a rising signal transition on the output of OR2, which is the primary output f0. All the signal transitions monotonically increase throughout the DIMS circuit, shown in Figure 3, from the primary inputs through the intermediate outputs to the primary outputs. Hence, the arrival of valid data on the primary inputs are said to be properly acknowledged by the primary outputs through the intermediate outputs. Although C1, C2, C4, C5, C6, and C7 also experience rising signal transitions on their inputs, they do not pose any problem of wire orphans [19] due to the isochronicity assumption imposed on the primary input forks.

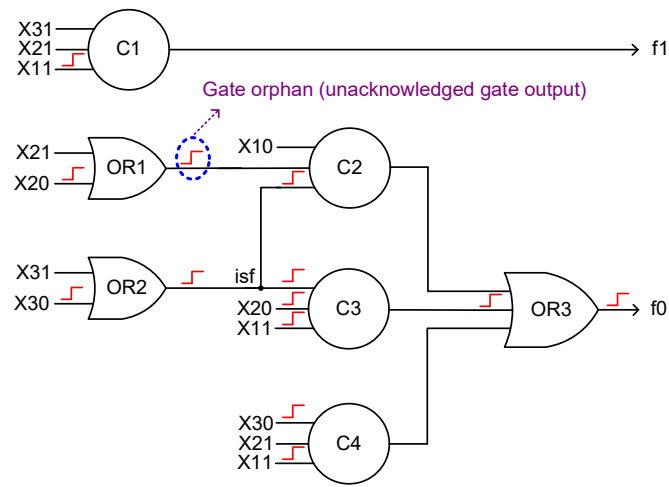

Figure 4. 3-input AND function synthesized based on the FDIMS approach [4]

When visually comparing Figures 3 and 4, it is evident that the 3-input AND function synthesized based on [4] is more optimized compared to that synthesized using the original DIMS method [5] since Figure 4 has less number of gates than Figure 3. However, Figure 4 tends to give rise to gate orphan(s) which is not present in Figure 3. This eventually undermines the robustness of [4]. To discuss this, let us consider the application of an example valid input data say, X30 = X20 = X11 = 1. For this input data, the respective rising signal transitions on X30, X20 and X11, and the corresponding rising signal transitions on the intermediate and primary outputs are shown in red in Figures 3 and 4. Similarly, the rising signal transitions for the application of this input data will be portrayed in red in Figure 5 which would follow.

In Figure 4, for the application of the input data X30 = X20 = X11 = 1, the output of OR2 experiences a rising signal transition which is followed by a rising signal transition on the output of C3 which is further followed by a rising signal transition on the output of OR3, which is the primary output f0. However, the output of OR1 also experiences a rising signal transition that would not be followed by any subsequent rising signal transition on the output of C2. In other words, the output of C2 would not acknowledge the risen signal transition on the output of OR1 since X10 = 0. As a result, the unacknowledged (risen) signal transition on the output of OR1, which is shown encompassed within the blue oval in dotted lines in Figure 4 is said to be a gate orphan that could be detrimental to the robustness of the asynchronous circuit shown in Figure 4, which is synthesized based on [4]. The reason for the occurrence of a gate orphan in Figure 4 is that the logic factorization suggested in [4] violates the safe QDI logic decomposition principles. No such gate orphan occurs in the DIMS circuit shown in Figure 3. This shall be clarified later through Table 1.

To overcome the gate orphan problem in Figure 4, a simple solution is suggested based on safe QDI logic decomposition which has been articulated through set theory principles in [39]. Accordingly, (2) is decomposed as follows. Since X30 and X20 are shared between the 1st and 2nd product terms on the right-hand side of (2), and either X10 or X11 would be asserted high during a valid data phase, these common literals can be extracted. Based on a similar technique, X30 and X21 can be extracted from the 3rd and 4th product terms, and X31 and X20 can also be extracted from the 5th and 6th product terms in (2). The resulting decomposed logic equation for f0 is given as,

$$f0 = X30X20(X10 + X11) + X30X21(X10 + X11) + X31X20(X10 + X11) + X31X21X10 \quad (4)$$

Figure 5 shows the synthesized 3-input AND function based on (1) and (4), and (4) is derived based on a safe QDI decomposition of (2) [14]. In Figure 5, nodes k1, k2 and k3 are isochronic forks; C1 to C11 represent the C-elements, with C2 to C4 and C6 to C11 being 2-input C-elements and C1 and C5 being 3-input C-elements. OR1 and OR2 are the OR gates. With reference to Figure 5, for the same input data applied viz. X30 = X20 = X11 = 1, we find that the output of C2 experiences a rising signal transition followed by the outputs of C7, OR1, and OR2 experiencing similar rising signal transitions. Notice that the risen signal transition on an input of C6 is said to be acknowledged since the risen signal transition on the other end of the isochronic fork k1, which is input to C7, is acknowledged through a rising signal transition on the output of C7. No gate orphan occurs in the circuit shown in Figure 5, which is synthesized through a safe QDI logic decomposition of the original DIMS solution.

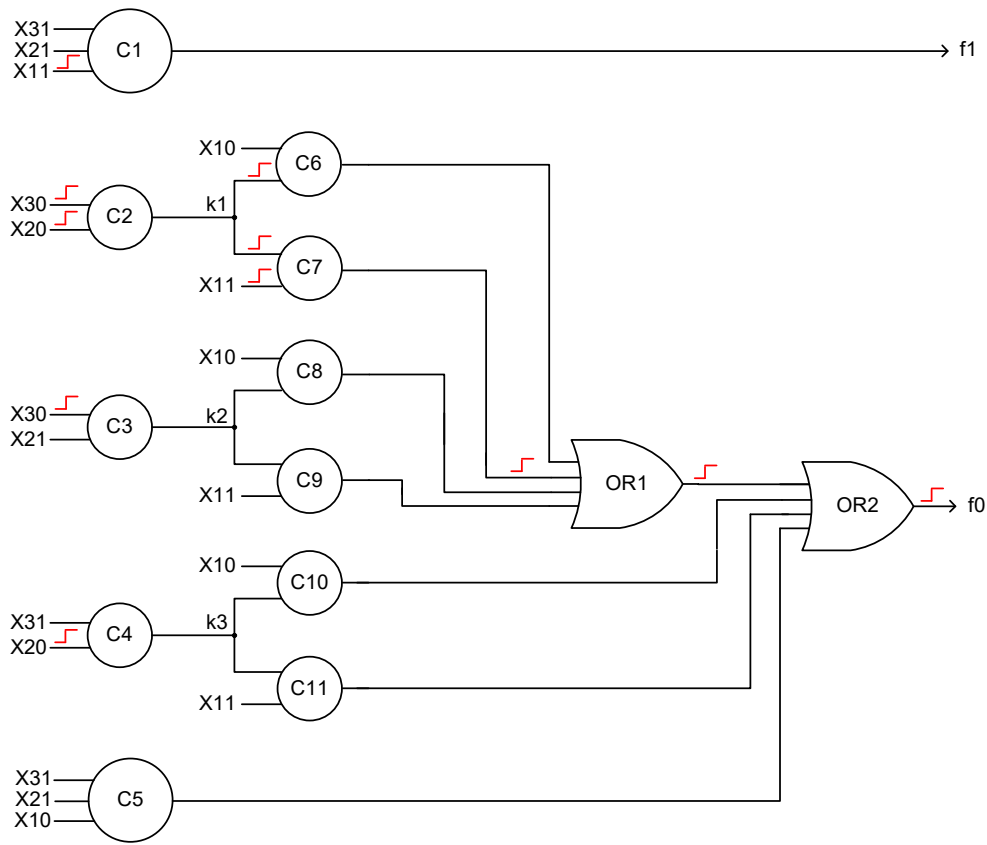

Figure 5. A gate orphan free asynchronous implementation of the 3-input AND function based on a safe QDI logic decomposition of the DIMS solution

Table 1 captures all the distinct input combinations corresponding to the dual-rail 3-input AND function, and shows the unique signal paths activated from the primary inputs to the primary output for the application of all the distinct input data to the different logic implementations shown in Figures 3, 4 and 5. The input combinations for which no gate orphan occurs or any gate orphan(s) occur in an asynchronous logic implementation are mentioned in Table 1. The sources of gate orphan(s) in Figure 4, corresponding to [4], are highlighted in italics.

Table 1. Application of distinct valid inputs to the different asynchronous implementations of the 3-input AND function shown in Figures 3, 4 and 5, and the mention of unique signal paths activated from the primary inputs to the primary output for the applied inputs

| Dual-rail primary inputs | | | | | | Figure 3 [5] | Figure 4 [4] | Figure 5 [14] |
| --- | --- | --- | --- | --- | --- | --- | --- | --- |
| X31 | X30 | X21 | X20 | X11 | X10 | | | |
| 0 | 1 | 0 | 1 | 0 | 1 | C2-OR1-OR2 | OR1-C2-OR3 | C2-OR1-OR7-OR8 |
| 0 | 1 | 0 | 1 | 1 | 0 | C3-OR1-OR2 | OR2-C3-OR3 *(Orphan due to OR1↑)* | C2-OR2-OR7-OR8 |
| 0 | 1 | 1 | 0 | 0 | 1 | C4-OR1-OR2 | OR1-C2-OR3 | C3-OR3-OR7-OR8 |
| 0 | 1 | 1 | 0 | 1 | 0 | C5-OR1-OR2 | C4-OR3 *(Orphans due to OR1↑, OR2↑)* | C3-OR4-OR7-OR8 |
| 1 | 0 | 0 | 1 | 0 | 1 | C6-OR1-OR2 | OR1-C2-OR3 | C4-OR5-OR7-OR8 |
| 1 | 0 | 0 | 1 | 1 | 0 | C7-OR1-OR2 | OR2-C3-OR3 *(Orphan due to OR1↑)* | C4-OR6-OR7-OR8 |
| 1 | 0 | 1 | 0 | 0 | 1 | C8-OR1-OR2 | OR1-C2-OR3 | C5-OR8 |
| 1 | 0 | 1 | 0 | 1 | 0 | C1-OR1-OR2 | C1 *(Orphans due to OR1↑, OR2↑)* | C1 |

In Table 1, the rising signal transition on a gate output is represented by the symbol ↑ that is placed next to the corresponding gate. The strongly indicating realization of the 3-input AND function shown in Figure 4, which is based on [4], may exacerbate the gate orphan(s) problem for certain input patterns, as seen in Table 1. It can be seen from Table 1 that no gate orphan occurs with respect to Figures 3 and 5 for all the input data, and gate orphan(s) tend to occur only in the case of Figure 4, which corresponds to [4]. In Table 1, we can see that when the data input to Figure 4 is X30 = X20 = X11 = 1 or X31 = X20 = X11 = 1, a gate orphan tends to occur due to an unacknowledged rising signal transition on the output of OR1. On the other hand, when the data input to Figure 4 is either X30 = X21 = X11 = 1 or X31 = X21 = X11 = 1, multiple gate orphans tend to occur due to unacknowledged rising signal transitions on the outputs of gates OR1 and OR2. Hence, for half of the input data patterns applied to Figure 4, gate orphan(s) tend to occur which does not augur well for the circuit robustness. Given this, the synthesis method of [4] is erroneously labelled as strongly indicating since indication in the intermediate output(s) is not complete. In fact, the potentially unsafe logic synthesis approach of [4] contradicts the tenets of indicating (i.e. QDI) asynchronous circuit design.

## 4. Confusion in the Synthesis Cost Estimation in [4]

It is to be noted that the synthesis cost of the DIMS method is given by the product terms comprising the logic expressions. Referring to (1) and (2), which are DIMS equations, the true rail of the 3-input AND function consists of 1 product term, and the false rail consists of 7 product terms. Thus, the synthesis cost of the original DIMS solution [5] is 8 product terms. On the other hand, the synthesis cost of (3) cannot be solely expressed as the number of product terms since (3) is a complex sum of products[2]. Although (3), based on [4], is expressed as a sum of products but internally it contains a product of sums. Therefore, the estimation of the synthesis cost of (3) is not straightforward unlike (2).

The author of [4] seems to have erratically estimated the cost of his FDIMS solution for some combinational logic benchmarks by supposedly enumerating only the product terms present in the final factorized expressions and discounted the presence of the product of sums present internally. This observation is vindicated by the fact that the author has provided an estimate of synthesis costs corresponding to some combinational benchmarks expressly in terms of the number of product terms alone – please see column 6 of Table 1 under Section 5 in [4]; this is incorrect. For example, with respect to the 3-input AND function considered, based on (1) and (3), the author of [4] has underestimated the cost of his FDIMS solution as 4 product terms i.e. 1 product term with respect to the true rail of the output, and 3 product terms with respect to the false rail of the output by neglecting the complex sum of products nature of (3), and eventually stating that the cost of the original DIMS solution which is 8 product terms has been reduced by 50% to just 4 product terms through logic factoring. This is an incorrect inference, which is substantiated by the author's statement given in lines 5 to 9 of the last paragraph of Section 5 of [4]. Therefore, due to the confusion arising from the incorrect estimation of the synthesis cost, it is suspected whether the results given in Table 1 of [4] are correct and reliable. Further, note that there is no generalized synthesis algorithm given for the FDIMS approach of [4], and this is yet another major drawback. The logic factorization performed in [4] is unsafe since it could give rise to gate orphan(s). This was articulated in Section 3 through the 3-input AND function that has been considered for illustration in [4].

Reference [4] considered only simple combinational benchmarks for experimentation which have less number of primary inputs. There is in fact a scalability issue with [4], which was not discussed and surreptitiously avoided. The original DIMS method [5] requires the enumeration of $2^n$ product terms for an $n$-input logic function. Consequently, it would be difficult to imagine how [4] could even attempt to factorize the original DIMS solution when $n$ may exceed 20 or 30 inputs since an input-space explosion is imminent for higher powers of 2 especially when $n$ becomes high [37]. Perhaps due to this underlying reason, [4] has considered only simple combinational benchmarks and has avoided the medium or bigger size benchmarks. Hence, [4] is not scalable and unsuitable for synthesizing even medium-size combinational logic functions leave alone the bigger benchmarks. This scalability issue is inherent in [6] and [7] as well. Therefore, for the efficient synthesis of arbitrary combinational logic functions as safe QDI asynchronous circuits, the methods presented in [16 – 20] are preferable as they build upon the foundation of a synchronous logic synthesis. Therefore, [4] has not advanced any existing knowledge in the field of indicating (i.e. QDI) asynchronous circuit synthesis but on the contrary, has caused confusions by discussing a sub-optimum, non-scalable, and an erroneous logic synthesis approach. Moreover, since [4] is additionally beset with the gate orphan problem, it is unsuitable for synthesizing strong-indication (QDI) asynchronous circuits.

---

[2] For example, Z = AB + CD is a simple sum of products, whereas G = (A + B) (C + D') + AD (B + E) is a complex sum of products.

## 5. Conclusions

This paper has critically commented on [4] by shedding light on the major problematic issues. The gate orphan problem imminent in the FDIMS approach of [4] was explained through an example 3-input AND function that was considered for illustration in [4]. In this backdrop, how to perform safe QDI logic decomposition was illustrated. The absence of a general logic synthesis algorithm in [4] was pointed out, and the confusion in the synthesis cost estimation were described. The input state-space explosion problem that is implicit in [4], which would adversely affect its scalability, was highlighted. It is beyond the scope of this critique to suggest a modification to the problem-ridden strongly indicating asynchronous logic synthesis approach of [4] as that would constitute a new research. Further, this is not deemed necessary given that better works such as [16 – 20] exist in the literature. These literature references are not only scalable but also efficient. Since the FDIMS approach of [4] is *merely said* to belong to the strong-indication category but is beset with the gate orphan problem and suffers from the problem of input-space explosion, it is risky to give it any consideration.

Further, it may be noted that in recent times, robust (i.e. safe) early output asynchronous logic synthesis which satisfies the QDI property, has assumed more importance in the realm of asynchronous circuit design. Robust early output logic is especially useful for arithmetic circuit designs which are basically iterative in nature and which require less latency, cycle time, area, and average power dissipation. A demonstration of these features can be found in [40] [41] with respect to 4-phase return-to-zero or return-to-one handshaking. Also, two things may be noted with respect to the design of asynchronous arithmetic circuits such as adders: i) strong-indication is not usually preferred since it always encounters the worst-case latencies and cycle time [42] [43], requires relatively more area, and dissipates relatively higher average power compared to the rest, and ii) weak-indication and early output adders are preferable, and among these the early output adders are more preferable from the perspectives of latencies, cycle time, area occupancy, and average power dissipation. These two observations have been confirmed in several publications on asynchronous adders with/without redundant logic [44] when targeting accurate and approximate computing by considering diverse architectures such as ripple carry [45 – 56], carry lookahead [57 – 60], and carry select [61] while employing homogeneous or heterogeneous delay-insensitive data encodings and following 4-phase return-to-zero or 4-phase return-to-one handshaking. However, for the design of datapath components such as multiplexers and demultiplexers, strong-indication may be preferred although it could result in more area, power and delay. For example, for the robust design of asynchronous carry select adders in [61], strongly indicating 2:1 multiplexers based on [62] are used to prevent any occurrence of gate orphans.